\documentclass[aps,preprint]{revtex4}
\usepackage{amsfonts}
\usepackage{amsmath}
\usepackage{amssymb}
\usepackage{graphicx}

\begin{document}

\title{Laser nanotraps and nanotweezers for cold atoms: 3D gradient dipole
force trap in the vicinity of Scanning Near-field Optical Microscope tip}
\author{V.V.Klimov}
\affiliation{P.N.Lebedev Physical Institute, Russian Academy of
Sciences, 53 Leninskii Prospect, 119991 Moscow, Russia} \email{
vklim@sci.lebedev.ru}
\author{S. K. Sekatskii and G. Dietler}
\affiliation{Laboratoire de Physique de la Mati\`{e}re Vivante,
IPMC, BSP, Ecole Polytechnique F\'{e}d\'{e}rale de Lausanne, CH
1015 Lausanne-Dorigny, Switzerland}
\begin{abstract}

Using a two-dipole model of an optical near-field of Scanning
Near-field Optical Microscope tip, i. e. taking into account
contributions of magnetic and electric dipoles, we propose and
analyze a new type of 3D optical nanotrap found for certain
relations between electric and magnetic dipoles. Electric field
attains a minimum value in vacuum in the vicinity of the tip and
hence such a trap is quite suitable for manipulations with cold
atoms.

\end{abstract}

\pacs{42.50.-p, 32.50.+d}
\maketitle

Recent enormous progress in the study of laser-cooled atoms and
molecules put forward the problem of their using for quantum
computing, frequency standard construction, and other
technological applications (see e.g. [1-3]). To achieve this goal,
as well as for the further progress of fundamental experiments in
the field, new compact traps and new methods of ``handling'' the
cold atoms, for example their transportation to/from a trap or
between different traps, should be elaborated. An example of
successful work in this direction is given e.g. by ``atom-chip
technology'' experiments  [4, 5], where guiding of cold atoms
along the wires  has been demonstrated. Further miniaturization of
such devices is desirable. Ideally, one should have at his/her
disposal true ``single cold atom nanotweezers'' which have very
small sizes and only slightly perturb the trap. Evidently, the
near-field optical configurations, based on subwavelength-size
aperture in the apex of a sharp fiber tip or local field
enhancement in the vicinity of sharp conducting tips (see e.g. [6,
7] for review on near-field optics) look rather promising. Hence
it is not surprising that a number of near-field optical
traps/tweezers have been proposed [7 - 15].

 However, all of them have certain drawbacks and, we believe, this is the main
reason why, to the best of our knowledge, none has been realized
up to now. Putting aside the configurations where an extremum of
the optical field is achieved on the surface of the tip [13]
(configuration obviously inappropriate for atom trap), among these
drawbacks we could mention that the traps proposed needed
different hard-to-control non-optical interactions (centripetal
potential, gravitational interaction, van der Waals force, etc.)
to be closed, can be realized only in the non practical light
reflection mode from the subwavelength aperture, and so on.

Here we propose and analyze the true ``free standing'', or
``support-free'' purely optical 3D trap emerging in vacuum in the
vicinity of an aperture of the Scanning Near-field Optical
Microscope (SNOM) tip. The characteristic size of our trap is
small in comparison with laser wavelength , so one indeed   can
speak about  nanotrap. Because methods of Angstrom-precision
motion of SNOM tip are well elaborated, the same construction is
cold atom nanotweezers.

\begin{figure}
\includegraphics[width=7in]{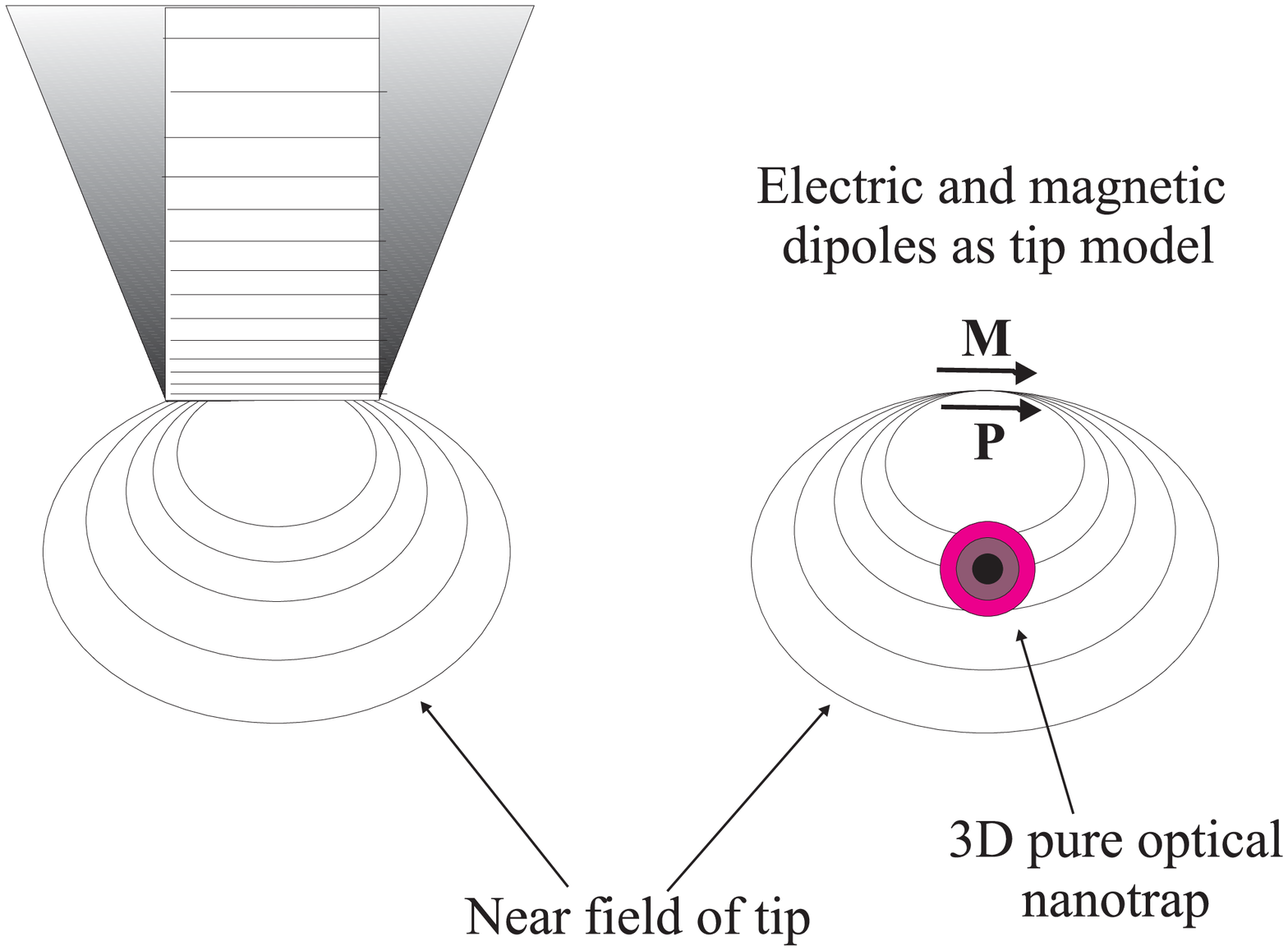}%
\caption{\label{fig1} Problem geometry  }
\end{figure}

Our analysis is based on the two-dipole model of optical near
field occurring in the vicinity of this tip, see Fig. \ref{fig1}.
Such a model has been established recently, when it has been shown
that in addition to the classic Bethe and Bouwkamp consideration,
where in the case of normal incidence optical near field is
modeled by one magnetic dipole \textbf{\textit{M}} (see e.g. Refs.
[16-18]), a field of an electric dipole \textbf{\textit{P}} should
be added to describe correctly optical near-field of a real fiber
tip \cite{Drezet1}-\cite{Drezet2}. This effect is due mainly to
the conical shape of the end area of such tip, and for a variety
of tips and light polarizations different relations between the
values and mutual orientations of
\textbf{\textit{M}},\textbf{\textit{ P}} dipoles can be
anticipated.

In near field of tip as in any laser fields, resonant atoms are
subject to the optical dipole force with the potential
\cite{Ashkin},\cite{Letokhov}:

\begin{equation}
\label{eq1}
U = {\frac{{\hbar \Omega}} {{2}}}\ln (1 + {\frac{{\mu ^{2}E_{las}^{2}
}}{{\hbar ^{2}\gamma ^{2}}}}) \approx {\frac{{\hbar \Omega
}}{{2}}}{\frac{{\mu ^{2}E_{las}^{2}}} {{\hbar ^{2}\gamma ^{2}}}}
\end{equation}

Here \textit{$\mu $} is an atomic transition dipole moment,
$\gamma ^{2} = \Omega ^{2} + (\Gamma / 2)^{2}$ where $\Gamma $ is
a natural line width, and $\Omega = \omega - \omega _{0} $ is
detuning between laser frequency $\omega $ and resonant frequency
of an atom $\omega _{{\rm 0}}$.  Throughout the paper we consider
the blue detuning $\Omega > 0$, which results in atom trapping at
the minimum of laser field intensity.It is well known that
blue-detuned gradient force optical traps based on minimum of an
electric field have essential advantages (they do not heat trapped
atoms, etc.) in comparison with those based on maximum of an
electric field using red-detuned light \cite{Letokhov}.

To characterize the proposed 3D nanotrap let us consider electric field of
light in the vicinity of a SNOM tip (i. e. in the near-field regions of both
dipoles). Corresponding electric field of an electric dipole \textbf{P} has
the form:

\begin{equation}
\label{eq2}
{\rm {\bf E}}_{e} = - {\frac{{{\rm {\bf P}}}}{{R^{3}}}} + 3{\frac{{{\rm {\bf
R}}\left( {{\rm {\bf R}}{\rm {\bf P}}} \right)}}{{R^{5}}}}
\end{equation}

\noindent
where \textbf{\textit{R}} is the radius vector from the dipole position to
an observation point. (SGCE units are used throughout the paper). Vector
potential of a magnetic dipole \textbf{M} has the following form

\begin{equation}
\label{eq3}
{\rm {\bf A}} = {\frac{{{\left[ {{\rm {\bf M}}{\rm {\bf R}}}
\right]}}}{{R^{3}}}}
\end{equation}

\noindent
and according to Faraday's law electric field of this magnetic dipole in the
near-field region has the form

\begin{equation}
\label{eq4}
{\rm {\bf E}}_{m} = ik{\frac{{{\left[ {{\rm {\bf M}}{\rm {\bf R}}}
\right]}}}{{R^{3}}}}
\end{equation}

\noindent where $k = {\dfrac{{\omega}} {{c}}}$ is wavevector in
free space. When two dipoles are located at the same point
\{0,0,0\}, intensity of total electric field ${\rm {\bf E}} = {\rm
{\bf E}}_{m} + {\rm {\bf E}}_{e} $ behind an aperture can be
presented in the form

\begin{equation}
\label{eq5}
\begin{array}{l}
 E^{2} = {\left| {{\rm {\bf E}}_{e} + {\rm {\bf E}}_{m}}  \right|}^{2} =\\
\\
 {\dfrac{{k^{2}{\left| {{\rm {\bf M}}_{}}  \right|}^{2}R^{2} - k^{2}\left(
{{\rm {\bf M}}{\rm {\bf R}}} \right)\left( {{\rm {\bf M}}_{}^{\ast}  {\rm
{\bf R}}} \right) + {\left| {{\rm {\bf P}}_{}}  \right|}^{2} + 3\left( {{\rm
{\bf P}}{\rm {\bf R}}} \right)\left( {{\rm {\bf P}}_{}^{\ast}  {\rm {\bf
R}}} \right) / R^{2} - ik{\rm {\bf R}}\left( {{\left[ {{\rm {\bf
P}}_{}^{\ast}  {\rm {\bf M}}} \right]} - {\left[ {{\rm {\bf P}}{\rm {\bf
M}}_{}^{\ast}}   \right]}} \right)}}{{R^{6}}}} \\
 \end{array}
\end{equation}

Below we will speak about this value as about an ``intensity'' having in
mind its obvious connection with the intensity of laser light $I_{las}$ ``seeping''
through the aperture of a SNOM tip: $I_{las} = {\dfrac{{c}}{{8\pi}} }E^{2}$,
where $c$ is speed of light .

It is naturally to look first for an axially symmetric trap. Such a trap
occurs in particularly if

\begin{equation}
\label{eq6}
\begin{array}{l}
 M_{z} = P_{z} = 0 \\
 M_{x} M_{y}^{\ast}  + M_{y} M_{x}^{\ast}  = 0 \\
 {\left| {M_{x}}  \right|} = {\left| {M_{y}}  \right|} = M_{0} \\
 P_{x} P_{y}^{\ast}  + P_{y} P_{x}^{\ast}  = 0 \\
 {\left| {P_{x}}  \right|} = {\left| {P_{y}}  \right|} = P_{0} \\
 \end{array}
\end{equation}

If these conditions hold true, expression for the intensity can be rewritten
in the form

\begin{equation}
\label{eq7}
E_{}^{2} = {\frac{{2k^{2}M_{0}^{2} R^{2} - k^{2}M_{0}^{2} \rho ^{2} +
2P_{0}^{2} + 3P_{0}^{2} \rho ^{2} / R^{2} - zkQ}}{{R^{6}}}}
\end{equation}

\noindent
where $Q = i\left( {{\left[ {{\rm {\bf P}}^{\ast} {\rm {\bf M}}}
\right]}_{z} - {\left[ {{\rm {\bf P}}{\rm {\bf M}}^{\ast}}  \right]}_{z}}
\right)$ and $\rho = \sqrt {x^{2} + y^{2}} $ is the distance from the
symmetry axis to an observation point.

Analysis shows that if we put the following additional condition on the
dipole momenta

\begin{equation}
\label{eq8}
Q^{2} = 16M_{0}^{2} P_{0}^{2}
\end{equation}

\noindent
intensity (\ref{eq7}) becomes equal to zero at the points

\begin{equation}
\label{eq9}
\left( {x = y = 0,kz = {\frac{{P_{0}}} {{M_{0}}} }} \right)  {\rm
}{\rm o}{\rm r} \left( {x = y = 0, kz = - {\frac{{P_{0}}} {{M_{0}
}}}} \right)
\end{equation}

In its turn, it is possible to show that (\ref{eq8}) will be satisfied in the case

\begin{equation}
\label{eq10}
\begin{array}{l}
 {\rm {\bf P}} = P_{0} {\left\{ {i,1,0} \right\}}, \\
 {\rm {\bf M}} = \pm M_{0} {\left\{ {i,1,0} \right\}} \\
 \end{array}
\end{equation}

It means that dipole momenta should be collinear to ensure the 3D trap.

\begin{figure}
\includegraphics[height=5in]{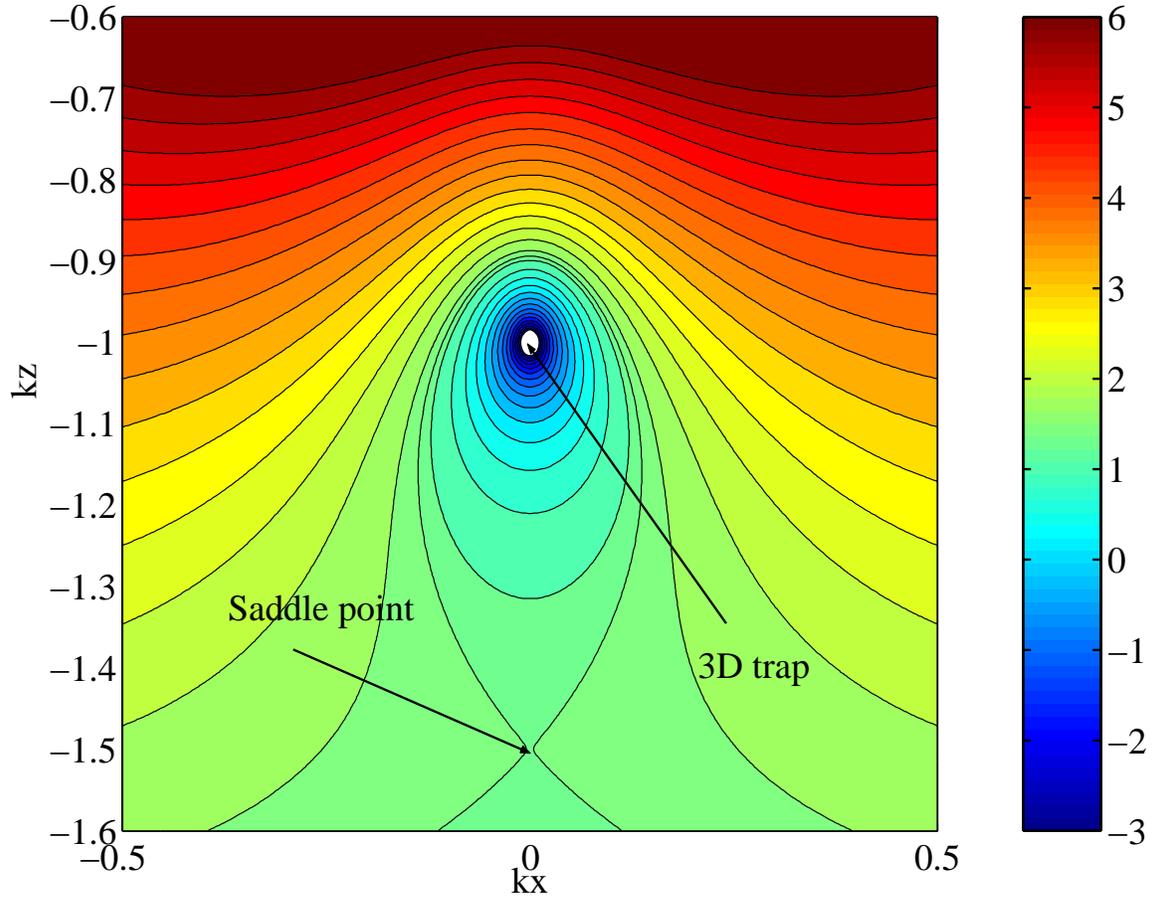}%
\caption{\label{fig2}   Isolines of the intensity eq. (\ref{eq7})
in the plane $y=0$ for $P_{0} = 10;M_{0} = 10$ (natural
logarithmic scale).}

\end{figure}

As intensity is a positive function of coordinates, zeros of
intensity (\ref{eq9})correspond to true 3D minimum of an electric
field. In Fig.\ref{fig2} the distribution of intensity in $x-z$
plane is shown for the case $P_{0} = 20;M_{0} = 10$.

Intensity (\ref{eq7}) has another point of extremum

\begin{equation}
\label{eq11}
\left( {x = 0,y = 0,kz = {\frac{{3}}{{2}}}{\frac{{P_{0}}} {{M_{0}}} }}
\right)   {\rm o}{\rm r}  \left( {x = 0,y = 0,kz =
- {\frac{{3}}{{2}}}{\frac{{P_{0}}} {{M_{0}}} }} \right)
\end{equation}

\noindent
which is a saddle point (see Fig. 2.). The value of intensity at this point
is

\begin{equation}
\label{eq12}
E^{2} = {\frac{{32}}{{729}}}k^{6}{\frac{{M_{0}^{6}}} {{P_{0}^{4}}} }
\end{equation}

This quantity can be used to estimate the potential well depth. For such an
estimation one can use $M_{0} = P_{0} = E_{0} a^{3}$, where $E_{{\rm 0}{\rm
}}$is an amplitude of the incoming light wave in aperture plane and $a$ is the
radius of an aperture [16-19]. Substituting these values into (\ref{eq12}), we get
for the potential well depth

\begin{equation}
\label{eq13}
\Delta E^{2} \sim {\frac{{32}}{{729}}}\left( {ka} \right)^{6}E_{0}^{2}
\end{equation}

For modern tips $ka \sim 1/2$ and hence $\Delta E^{2} \sim 10^{-3}E_{0}^{2}$.
 For typical experimental conditions, an intensity of optical
near field at the aperture of SNOM tip is about $I_{0} = 10^{{\rm
3}} - 10^{{\rm 4}} W/cm^{{\rm 2}}$ [6,7]. This means that e. g.
for alkaline atoms, which are characterized by resonant dipole
moments of the order of 10$^{{\rm -} {\rm 1}{\rm 7}}$ SGSE and
natural line widths $\Gamma \cong 10^{7} - 10^{8} s^{- 1}$ (for
example, for the 6S$_{{\rm 1}{\rm /} {\rm 2}}$--6P$_{{\rm 3}{\rm
/} {\rm 2}}$ D2 transition of cesium atom at \textit{$\lambda
$}=852 nm, \textit{$\mu $}=8.01$\cdot $10$^{{\rm -} {\rm 1}{\rm
8}}$ CGSE and $\Gamma = 3.07 \cdot 10^{7} s^{- 1}$ \cite{Rafac}),
traps with the depth of the order of a few milliKelvin, what is
quite standard for optical dipole gradient force -- based traps,
can be realized when using $\Omega \cong 100\Gamma $, see
(\ref{eq1}). It is very important that under these conditions the
trap will have about 10 energy levels of atomic motion with lowest
level being about $10^{-4}K $.

\begin{figure}
\includegraphics[height=5in]{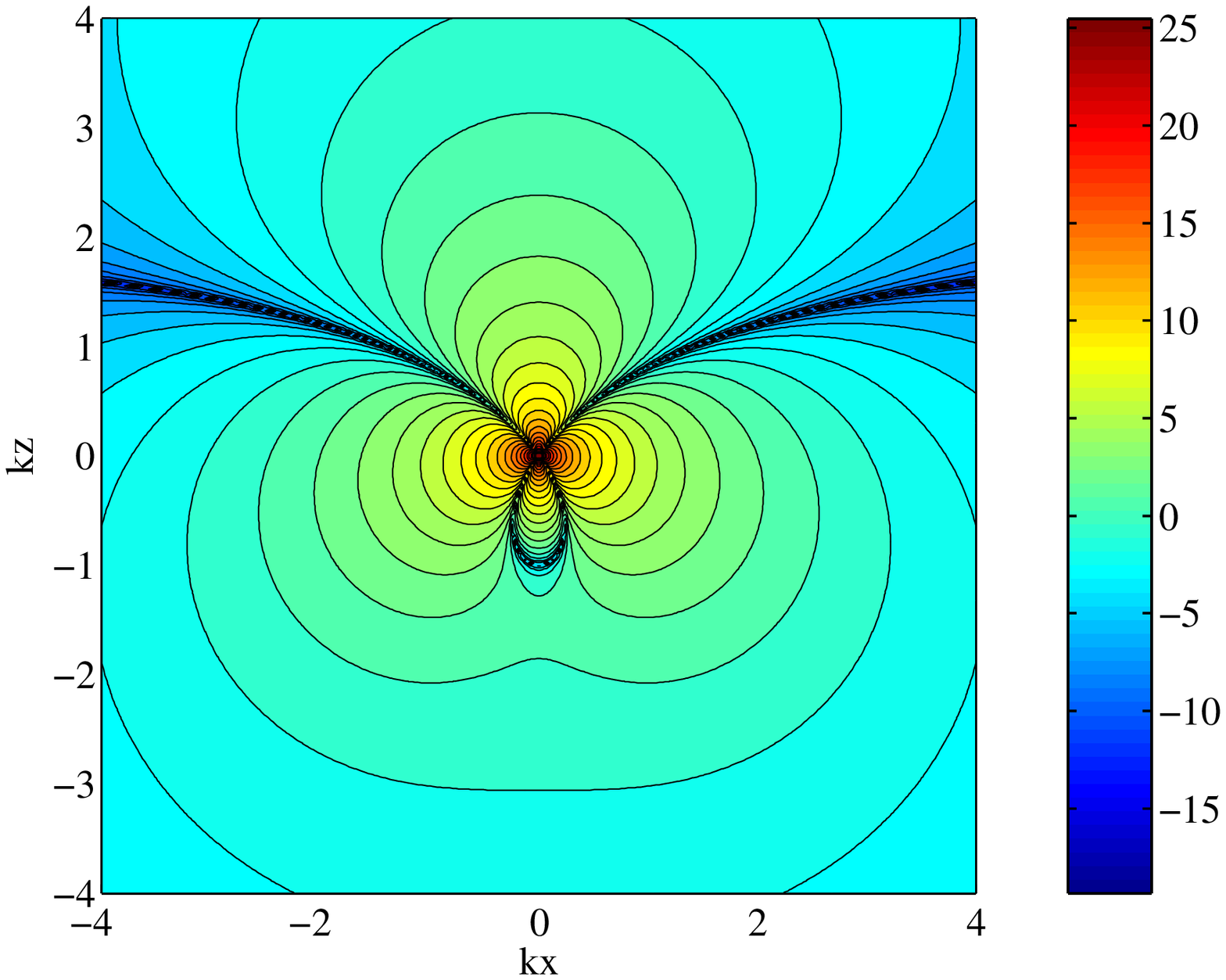}%
\caption{\label{fig3} Isolines of ${\left| {{\rm {\bf%
E}}_{x}}\right|}^{2}$ in the vicinity of the 3d trap in the plane
$y=0$ for $P_{0} = 10;M_{0} = 10$ (natural logarithmic scale). }

\end{figure}

\begin{figure}
\includegraphics[height=4in]{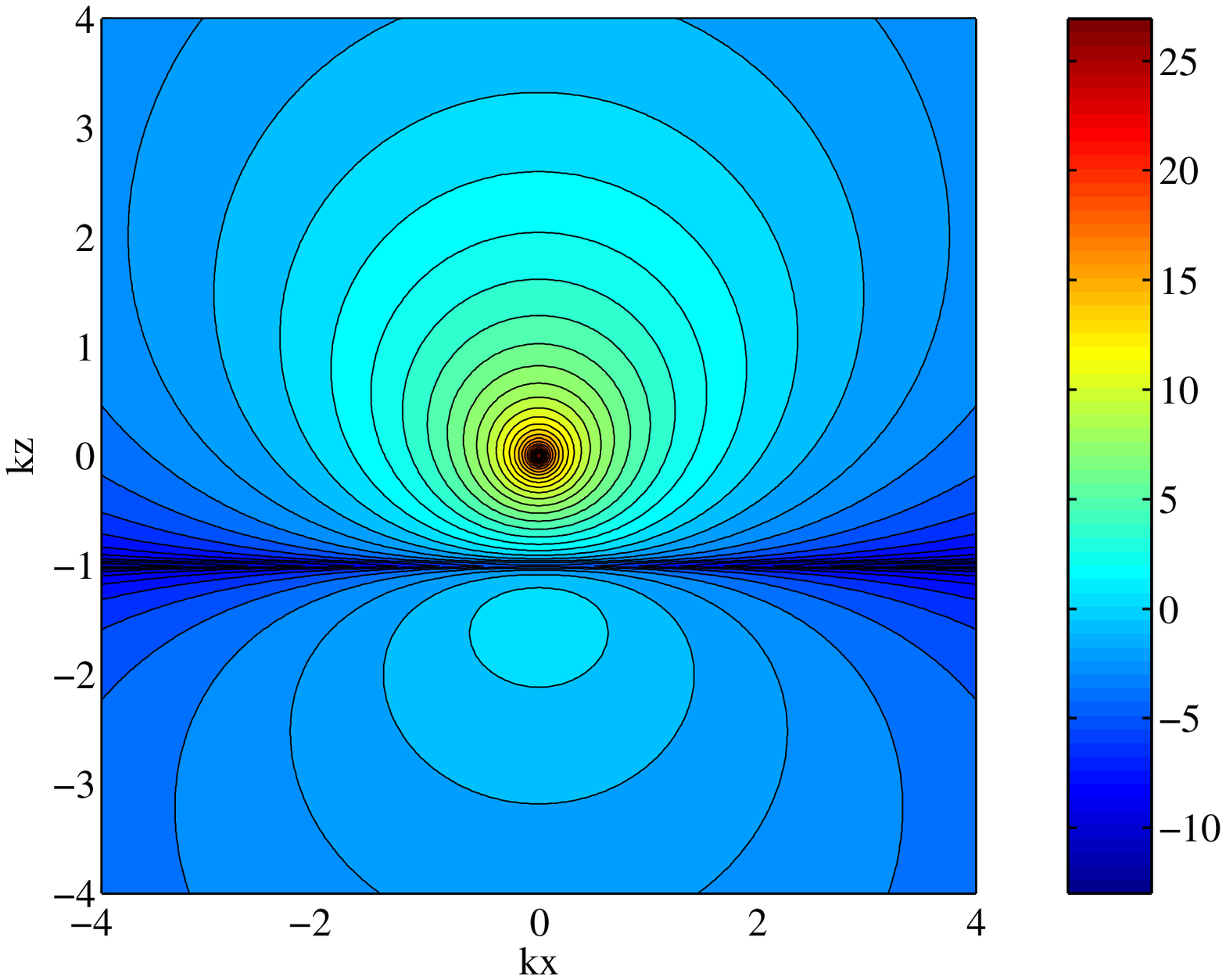}%
\caption{\label{fig4} Isolines of ${\left| {{\rm {\bf%
E}}_{y}}\right|}^{2}$ in the vicinity of the 3d trap in the plane
$y=0$ for $P_{0} = 10;M_{0} = 10$ (natural logarithmic scale). }
\end{figure}

\begin{figure}
\includegraphics[height=4in]{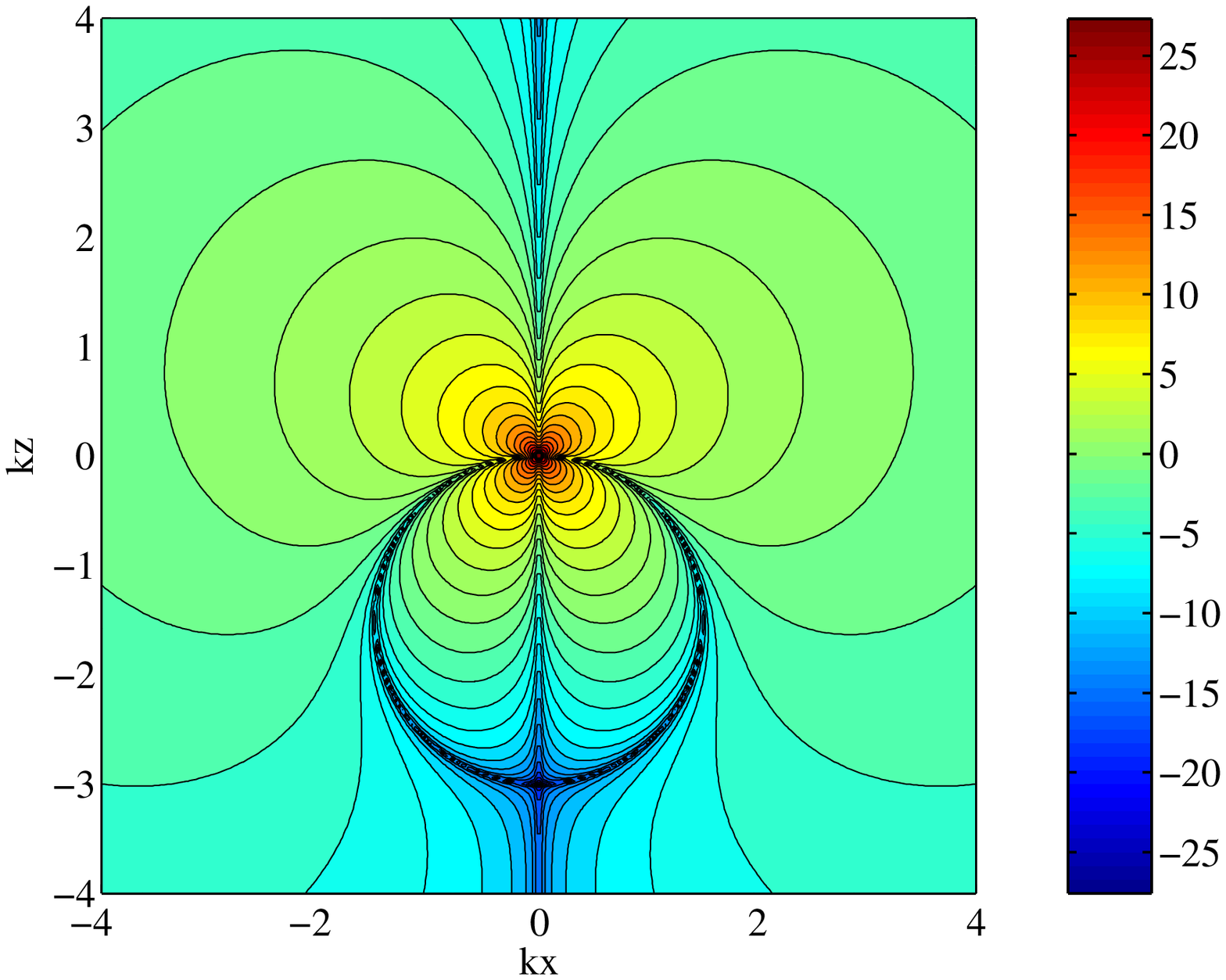}%
\caption{\label{fig5} Isolines of ${\left| {{\rm {\bf%
E}}_{z}}\right|}^{2}$ in the vicinity of the 3d trap in the plane
$y=0$ for $P_{0} = 10;M_{0} = 10$ (natural logarithmic scale). }
\end{figure}

It is worthwhile to note, that existence of such a 3D trap is
highly nontrivial, because the relevant field components have very
complicated structure (see Figs. \ref{fig3}, \ref{fig4},
\ref{fig5}), and it seems very difficult to provide a minimum for
their sum.

Mathematically, by varying \textbf{\textit{P}} and
\textbf{\textit{M}} ratio and other parameters of the problem, our
trap can be placed at any point on the symmetry axis. However, the
two dipole approximation used is not valid very close to the
aperture plane $z = 0$. Its validity starts from $z \sim a$, where
$a$ is the radius of an aperture. Hence we should consider only
such parameters where the trap position occurs not too close to
the aperture plane. Besides the condition $\mid z \mid  \gtrsim a$
allows us to neglect van der Waals attractive force which is
always important in close vicinity of tip surface.

\begin{figure}
\includegraphics[height=4in]{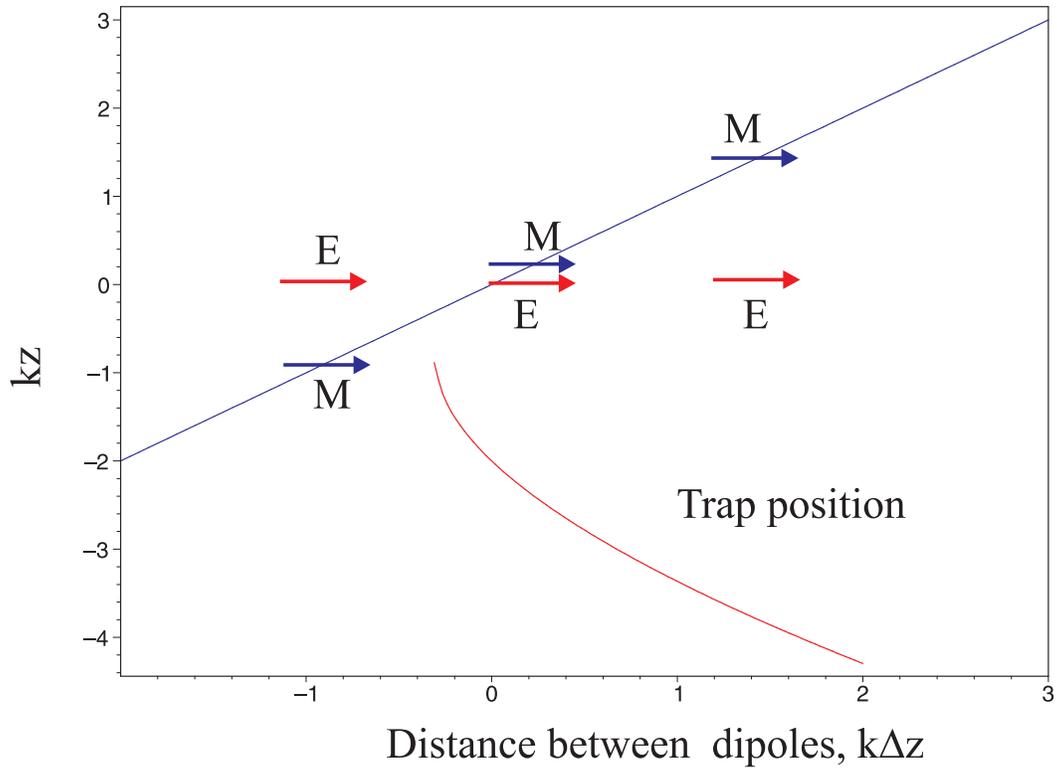}%
\caption{\label{fig6}   Variation of the trap position as a function of axial splitting between the
dipoles. }
\end{figure}

For our trap the minimum of intensity is stable against small
perturbations. For example, if relation between the dipole values
is

\begin{equation}
\label{eq14}
\begin{array}{l}
 {\rm {\bf P}} = P_{0} {\left\{ {i,1,0} \right\}}, \\
 {\rm {\bf M}} = \pm M_{0} e^{i\varphi} {\left\{ {i,1,0} \right\}} \\
 \end{array}
\end{equation}

\noindent then true 3D minimum still exists provided $\sin {\left|
{\varphi}  \right|} < {\frac{{1}}{{5}}}$ or $ - 0.2 < \varphi <
0.2$. Small variations of z-components of momenta ($P_{z}$ and
$M_{z})$ also result in small variations of trap position. Small
mutual displacements (splitting) of dipoles in axial and/or radial
directions also result in variation of the trap. In the case when
magnetic dipole is moving away from the trap, that is in the case
when electric dipole is placed between the trap and the magnetic
one, the trap suffers only minor shifts. In the case when magnetic
dipole is placed between an electric one and the trap, the
intensity minimum disappears for large enough splitting, see
Fig.\ref{fig6}.

The retardation effects in the near field region generally are
small. Nevertheless to estimate their influence let us consider
the full electric field of electric and magnetic dipoles:

\begin{equation}
\label{eq15}
\begin{array}{l}
 {\rm {\bf E}}_{e} = {\rm {\bf P}}\left( { - {\dfrac{{1}}{{R^{3}}}} +
{\dfrac{{ik}}{{R^{2}}}} + {\dfrac{{k^{2}}}{{R}}}} \right) + {\dfrac{{{\rm {\bf
R}}\left( {{\rm {\bf R}}{\rm {\bf P}}} \right)}}{{R^{2}}}}\left(
{{\dfrac{{3}}{{R^{3}}}} - {\dfrac{{3ik}}{{R^{2}}}} - {\dfrac{{k^{2}}}{{R}}}}
\right) \\
 {\rm {\bf E}}_{m} = {\dfrac{{{\left[ {{\rm {\bf M}}{\rm {\bf R}}}
\right]}}}{{R^{}}}}\left( {{\dfrac{{ik}}{{R^{2}}}} + {\dfrac{{k^{2}}}{{R}}}}
\right) \\
 \end{array}
\end{equation}

Analysis shows that for such a case 3d trap survives only when $\xi = P_{0}
/ M_{0} < 1.5035$. Positions of the trapping region and the saddle point are
given respectively by the formulae

\begin{equation}
\label{eq16} \left( {x = y = 0,kz = \xi - \xi ^{3} + 3\xi ^{5} +
...} \right) {\rm } {\rm o}{\rm r} \left( {x = y = 0, kz = -
\left( {\xi - \xi ^{3} + 3\xi ^{5}+...} \right)} \right)
\end{equation}

\noindent
and

\begin{equation}
\label{eq17}
\left( {x = y = 0,kz = {\frac{{3}}{{2}}}\xi - {\frac{{9}}{{16}}}\xi ^{3} -
{\frac{{189}}{{128}}}\xi ^{5} + ...} \right)   {\rm o}{\rm
r} \left( {x = y = 0, kz = - \left( {{\frac{{3}}{{2}}}\xi -
{\frac{{9}}{{16}}}\xi ^{3} - {\frac{{189}}{{128}}}\xi ^{5}...} \right)}
\right)
\end{equation}

Intensity at the bottom of the potential well is now nonzero

\begin{equation}
\label{eq18}
E_{\min} ^{2} = 2k^{6}M_{0}^{2} \left( {\xi ^{2} - \xi ^{4} + ...} \right)
\end{equation}

\noindent
while the intensity at the saddle point is

\begin{equation}
\label{eq19}
E_{saddle}^{2} = k^{6}M_{0}^{2} \left( {{\frac{{32}}{{729\xi ^{4}}}} +
{\frac{{40}}{{81\xi ^{2}}}} + ...} \right)
\end{equation}

These values determine the depth of our trap. Comparing results
(\ref{eq16})-(\ref{eq19}), where retardation effects are taken
into account, with the quasistatic results (\ref{eq11}) we see
that in the case of magnetic dipole domination $\left( {\xi
\lesssim 1} \right)$ the position of the trap remains in the near
field region and hence the retardation effects have only minor
influence. On the other hand, in the case of substantially large
amplitude of electric dipole, the retardation effects destroy our
trap.

Hence we have shown that minimum of an electric field with the
size smaller than the light wavelength and depth of a few
milliKelvin did occur in vacuum in the vicinity of a SNOM tip.
This minimum is stable against perturbations and can be tuned both
in position and depth by changing relation between $P_0$ and $M_0$
and incident light electric field amplitude E0. This attests the
trap proposed as a very promising basic element for future cold
atom nanotraps and nanotweezers. Finally, we would like to note
the following. Despite the two dipole model of optical near-field,
whose using is inherent to obtain the reported results, nowadays
seems is well established and supported by experiments [19-21],
complete and rigorous analysis of all possible dipole
configurations occurring for near field of a real tip is still
lacking. In particular, the conditions to be imposed on
experimental setup to obtain the trapping configuration of
magnetic and electric dipoles also remain to be understood.
Nevertheless, we believe that broad possibilities to vary SNOM tip
shapes and coatings [6, 7] as well as to vary other parameters (e.
g. incoming light polarization) give enough hope for practical
realization of such a configuration. Indeed, it can quite happen
that this is already done at least for one of plenty of the tips
demonstrated up to date.

\begin{acknowledgments}

The authors are grateful to Swiss National Science Foundation and
Russian Foundation for Basic Research  (V.V.K., grant \#
04-02-16211) for financial support of this work.

\end{acknowledgments}

\newpage

\newpage

\newpage

\newpage

\newpage

\newpage

\end{document}